\documentstyle[11pt,aaspp4]{article}
\newcommand{\dmm}{\mbox{$\Delta$m$_{15}(B)$}}
\newcommand{\kms}{\mbox{km s$^{-1}$}}
\newcommand{\ho}{\mbox{$H_\circ$}}
\begin{document}

\title{The Reddening-Free Decline Rate Versus Luminosity Relationship\\
for Type~Ia Supernovae}

\author{M. M. Phillips
}
\affil{Las Campanas Observatory, Carnegie Observatories, Casilla 601,
La Serena, Chile
}
%\authoremail{mmp@lco.cl}
%%%%
\and
\author{Paulina Lira
}
\affil{ Department of Physics \& Astronomy, University of Leicester, 
Leicester, LE1 7RH, U.K.
}
%\authoremail{plt@star.le.ac.uk}
%%%%
\and
\author{
Nicholas B. Suntzeff, R. A. Schommer
}
\affil{
Cerro Tololo Inter-American
Observatory, National Optical Astronomy
Observatories,\altaffilmark{1}\\ Casilla 603, La Serena, Chile 
}
%%%%
\and
\author{
Mario Hamuy
}
\affil{University of Arizona, Steward Observatory, Tucson, Arizona 85721
} 
%\authoremail{mhamuy@as.arizona.edu}
%%%%
\and
\author{
Jos\'{e} Maza\altaffilmark{2}
}
\affil{Departamento de Astronom\'{\i}a, Universidad de Chile, Casilla 36-D, 
Santiago, Chile
}
%\authoremail{jose@das.uchile.cl}
%%%%
\altaffiltext{1}{
Cerro Tololo Inter-American Observatory, National
Optical Astronomy Observatories, operated by the Association of Universities
for Research in Astronomy, Inc., (AURA), under cooperative agreement with
the National Science Foundation.
}
\altaffiltext{2}{
C\'{a}tedra Presidencial de
Ciencias (Chile) 1995
}

\begin{abstract}

We develop a method for estimating the host galaxy dust extinction for 
type~Ia supernovae based on an observational coincidence first noted by 
\cite{Lir95}, who found that the $B-V$ evolution during the period 
from 30--90 days after $V$ maximum is remarkably similar for all events, 
regardless of light curve shape. This fact is used to calibrate the 
dependence of the $B_{max}-V_{max}$ and $V_{max}-I_{max}$ colors on 
the light curve decline rate parameter \dmm, which can, in turn, be used to
separately estimate the host galaxy extinction.  Using these methods to 
eliminate the effects of reddening, we reexamine the functional form of the 
decline rate versus luminosity relationship and provide an updated 
estimate of the Hubble constant of 
\ho = 63.3 $\pm$ 2.2(internal) $\pm$ 3.5(external) \kms\ Mpc$^{-1}$.
\end{abstract}


\section{Introduction}

In recent years, considerable attention has been focused on the utility of 
type~Ia supernovae (SNe~Ia) as cosmological standard candles (e.g.,
\cite{Ham_etal96b,Per_etal95,Rie_etal96a,San_etal96}). 
While there is now abundant evidence for the existence of a 
significant dispersion in the peak luminosities of these events at optical 
wavelengths, the absolute magnitudes fortuitously appear to be closely 
correlated with the decay time of the light curve 
(\cite{Phi93,Ham_etal95,Rie_etal95,Ham_etal96a,Rie_etal96a}).  
After correction for 
this effect, dispersions of $\sim$0.15-0.20~mag were obtained by 
\cite{Ham_etal96b} in the $BVI$ Hubble diagrams of 26 SNe~Ia 
in the redshift 
range 0.01 $\leq z \leq$ 0.1, and a Hubble constant of 
63.1 $\pm$ 3.4 (internal) $\pm$ 2.9 (external) \kms\ Mpc$^{-1}$ 
was derived via reference to four nearby SNe~Ia with 
Cepheid-calibrated distances.  Although \cite{Ham_etal96b} 
applied a color cut to 
their sample to eliminate the reddest events, no corrections were applied 
for possible host galaxy dust extinction of the supernovae in either the
distant or nearby samples. Not only might 
this affect the derived slope of the decline rate versus luminosity relation, 
but the correctness of the Hubble constant depends on the assumption that 
the mean reddening of the 26 distant SNe~Ia in the Hubble flow is essentially 
the same as that of the four nearby calibrating SNe.

In a parallel study which included many of the objects observed by 
\cite{Ham_etal96b}, \cite{Rie_etal96a} developed an empirical method which they 
called ``Multicolor Light Curve Shapes'' (MLCS) to estimate simultaneously
the luminosity, distance, and total extinction of each SN~Ia.  MLCS employs 
linear estimation algorithms to create a family of SN~Ia light curves and 
color curves, and offers considerable elegance in error modeling.
However, this first implementation by \cite{Rie_etal96a} was based on 
essentially the same ``training set'' of only nine nearby supernovae
used by \cite{Phi93}
which suffered from uncertain extinction estimates and 
inhomogeneous secondary distance indicators.  This training set
was sufficient to determine the existence of the relations between
SN~Ia light curve shape, color, and luminosity, but not ideal for
determining the precise values of these relations. 
Indeed, a recent reanalysis of MLCS in the $B$ and $V$ bands by 
\cite{Rie_etal98} using a much larger training set as well as redshifts
as the distance indicator to determine the SN~Ia luminosities improves
the precision of the method.  Additionally, this reanalysis reduces the 
strong relation between light curve shape and $B-V$ color at maximum
light initially found by \cite{Phi93} and \cite{Rie_etal96a}.

In this paper, we seek to establish an independent procedure for determining 
host galaxy reddenings 
for SNe~Ia using an observational coincidence first noted 
by \cite{Lir95}.  This method is developed in \S~2 and used to estimate 
the intrinsic $B_{max}-V_{max}$ and $V_{max}-I_{max}$ colors 
for a sample of $\sim$60 well-observed SNe~Ia.  We then apply this 
information in \S~3 to derive reddening-free relations between the decline 
rate parameter \dmm\ (\cite{Phi93}) and the peak absolute magnitudes in 
$BVI$, and to update our estimate of the value of the Hubble 
constant.

\section{Color Excesses and Intrinsic Colors}

\subsection{$E(B-V)_{Tail}$}

In comparing the color evolution of several apparently unreddened SNe~Ia 
representing the full range of observed light curve decline rates, 
\cite{Lir95} found that the $B-V$ colors at 30--90 days past $V$ maximum 
evolved in a nearly identical fashion.  This property, which was 
independently discerned by \cite{Rie_etal96a}, is illustrated in Figure \ref{f1}
where we plot the $B-V$ color evolution from CTIO photometry of six SNe~Ia 
covering a wide range of decline rates (0.87 $\leq$ \dmm $\leq$ 1.93) and which 
likely suffered little or no reddening from dust in their host galaxies.  
The likelihood of a supernova being unreddened was judged by 
three basic criteria: 1) the absence of interstellar Na~I or Ca~II lines in 
moderate-resolution, high signal-to-noise spectra, 2) the morphology of the 
host galaxy (SNe that occur in E or S0 galaxies are less likely to be 
significantly affected by dust), and 3) the position of the supernova in the 
host galaxy (SNe lying outside the arms and disks of spirals are more likely 
to have low dust reddening).  As discussed by Lira, it is not surprising 
that the colors of SNe~Ia at these late epochs are so similar since their 
spectra also display an impressive uniformity.  This is precisely the epoch 
where the Fe-Ni-Co core dominates the spectrum which is rapidly evolving 
into the ``supernebula'' phase (see \cite{Phi_etal92}).

From a least squares fit to the photometry of four of the SNe displayed in 
Figure \ref{f1} (1992A, 1992bc, 1992bo, and 1994D), Lira derived the 
following relation to describe the intrinsic B-V color evolution:

\begin{equation}
(B-V)_\circ = 0.725 - 0.0118 (t_V - 60),  \label{eq:eq1}
\end{equation}

\noindent
where $t_V$ is the phase measured in days since $V$ maximum.  This fit, 
which is plotted in Figure \ref{f1}, is valid over the phase interval 
30 $\leq t_V \leq$ 90, and can be applied to any SN~Ia with $BV$ coverage 
extending to at least $t_V$ = 30 days to derive an estimate of the color 
excess $E(B-V)$.  Note in Figure \ref{f1} that individual SNe can display
systematic residuals with respect to the Lira fit (i.e., the points lie
mostly bluer or redder than the line) suggesting that there is an intrinsic
dispersion about equation~\ref{eq:eq1}.  This dispersion amounts to 0.04~mag
for the four SNe used by Lira, and 0.06~mag for the six SNe shown in 
Figure \ref{f1}; in calculating color excesses from equation~\ref{eq:eq1},
we shall adopt a value of 0.05~mag in our error analysis.

For relatively bright SNe observed with CCDs, the 
precision of the photometry obtained at $\sim$1--3 months after maximum is 
typically still quite good (0.02-0.05~mag).  However, for more distant 
events, the late-time photometry coverage tends to be relatively sparse 
with typical errors $\sim$0.1--0.2~mag for individual observations, and 
values of $E(B-V)$ derived directly from equation~\ref{eq:eq1} will often 
have large associated errors.  For such SNe, a better technique is 
to use template fits of the type employed by \cite{Ham_etal96c} to 
derive an estimate of the observed $B-V$ color 
at some fiducial epoch in the interval 30 $\leq t_V \leq$ 90.  This value 
can then be compared with the intrinsic color predicted by 
equation~\ref{eq:eq1} for the same epoch to derive the color excess.  
This procedure has the advantage of using the entire set of $BV$ photometry 
to estimate a late-epoch color rather than just the subset of data 
obtained at 30 $\leq t_V \leq$ 90.

For a sample of 62 well-observed SNe~Ia with $z \leq 0.1$ consisting of 1) the 
``Cal\'{a}n/Tololo'' sample of 29 SNe~Ia published by \cite{Ham_etal96c}, 
2) 20 SNe~Ia published in \cite{Rie_etal99} which we shall refer to as the
``CFA'' sample, and 
3) 13 nearby, well-observed SNe~Ia (1937C, 1972E, 1980N, 1981B, 1986G, 
1989B, 1990N, 1991T, 1991bg, 1992A, 1994D, 1996X, 
and 1996bu --- see Table \ref{t1}
for photometry references), we have calculated
color excesses from equation~\ref{eq:eq1} via the two techniques described
above --- i.e., both directly from the late-time photometry and via light
curve template fits.  In the following, we shall refer to these as 
the ``Direct'' and ``Template'' methods, respectively.  
For the ''Template'' method, we chose $t_V$ = 53 days as the fiducial 
reference epoch for which equation~\ref{eq:eq1} predicts 
$(B-V)_\circ = 0.808$.  To calculate the host galaxy reddening, 
$E(B-V)_{host}$, the observed supernova color evolution must be 
corrected for both Galactic reddening and the effect of redshift (the 
``K~correction'') via the equation 

\begin{equation}
(B-V)_{corr}=(B-V)_{obs}-E(B-V)_{Gal}-K_{B-V}, \label{eq:eq2}
\end{equation}

\noindent
where $(B-V)_{obs}$ is the observed color, $E(B-V)_{Gal}$ is the 
Galactic reddening, for which we use the values 
from \cite{Sch_etal98}, and $K_{B-V}$ is the $K$~correction.  The host 
galaxy reddening is then given by

\begin{equation}
E(B-V)_{host}=(B-V)_{corr}-(B-V)_\circ. \label{eq:eq3}
\end{equation}

As discussed by \cite{Rie_etal98} and \cite{Nug_etal99}, 
$K_{B-V}$ is not only a function of 
redshift but also the SN color.  Since we are concentrating on an epoch 
where all SNe~Ia show essentially the same $B-V$ color, we need only worry 
about color differences arising from different amounts of host galaxy 
reddening.  In principle, an iterative procedure should be followed where 
$K_{B-V}$ for zero host galaxy reddening (\cite{Ham_etal93}) is taken as a 
first guess, $E(B-V)_{host}$ is calculated via equations~\ref{eq:eq2} and 
\ref{eq:eq3}, and the K$_{B-V}$ correction is then adjusted if $E(B-V)_{host}$ 
is found to be significantly greater than zero.  However, for the range of 
host galaxy reddenings ($E(B-V) < 0.8$) typical of our sample, the 
variation of K$_{B-V}$ due to this effect is only 
0.00--0.02~mag. This is much smaller than the precision of our reddening 
indicator, and so we therefore adopt the K$_{B-V}$ corrections for zero 
extinction with no iteration.

As pointed out by \cite{Kim97}, the observed color excess of a SN~Ia will 
vary slightly with light curve phase due to the significant color evolution 
that these events undergo.  There will also be a generally smaller 
dependence on the total amount of dust extinction itself (e.g., see 
\cite{Bla56}).  Using the IRAF task ``deredden'', which employs a standard 
Galactic interstellar reddening curve (\cite{Car_etal89}), and a program 
which we have developed for deriving synthethic photometry from spectra, 
we have calculated observed color excesses as a function of both the light 
curve phase and the ``true'' reddening, $E(B-V)_{true}$, for a database of 
more than 30 SNe~Ia spectra in order to characterize and correct for these 
effects.  Note that $E(B-V)_{true}$ parameterizes the amount of extinction 
along a particular line of sight, and {\em not} the observed color excess 
which, as stated, will vary as a function of phase and reddening.  To convert
from $E(B-V)_{obs}$ to $E(B-V)_{true}$ at $t_V$ = 53 days, our reference
epoch for the ``tail'' evolution phase, we find the following approximate
relation to be valid:

\begin{equation}
E(B-V)_{true} = 1.018 / (1/E(B-V)_{obs} - 0.072). \label{eq:eq4}
\end{equation}

The observed decline rate of a SN~Ia will also be a weak function of
the dust extinction which affects the light curves.  This follows from 
Figure \ref{f2}, where we have used the above-described calculations to 
examine the variation of

\begin{equation}
R_B = A_B / E(B-V)_{true} \label{eq:eq5}
\end{equation}

\noindent
with light curve phase
for three different values of $E(B-V)_{true}$.  In this equation, $A_B$
is the {\em observed} extinction which is equal to the difference between
the synthetic $B$ magnitudes calculated for a reddened and unreddened 
spectrum.  Combining equation~\ref{eq:eq5} with the definition of the
observed decline rate parameter,

\begin{displaymath}
\dmm_{obs} = B_{obs}(+15~days) - B_{obs}(max),
\end{displaymath}

\noindent
gives

\begin{displaymath}
\dmm_{obs} = \dmm_{true}+[R_B(+15~days)-R_B(max)]E(B-V)_{true}
\end{displaymath}

\noindent
From Figure \ref{f2}, this can be shown to reduce to the 
following approximate relation:

\begin{equation}
\dmm_{true} \simeq \dmm_{obs}+0.1E(B-V)_{true} \label{eq:eq6}
\end{equation}

\noindent
Note that reddening acts to {\em decrease} the decline rate, which is 
opposite to the conclusion of \cite{Lei88} who employed a formula 
from \cite{Sch82} which gives $A_B$ as a function of the 
intrinsic color, $(B-V)_o$, and the observed color excess, $E(B-V)_{obs}$.
Although this formula was derived for normal stars, it does a surprisingly 
reasonable job of predicting the shape of the variations of $A_B$ with 
light curve phase implied by Figure \ref{f2}.  However, to use this 
formula correctly, one must take into account that $E(B-V)_{obs}$ 
also varies with phase; \cite{Lei88} assumed that $E(B-V)_{obs}$ 
was constant with phase, which led him to incorrectly conclude that the 
effect of reddening was to increase the decline rate.

Figure \ref{f3} illustrates application of the ``Direct'' and ``Template'' 
methods to the photometry of two representative SNe observed in the 
course of the Cal\'{a}n/Tololo survey.  Comparison of the 
$E(B-V)_{host}$ values (converted to ``true'' reddenings via
equation~\ref{eq:eq4}) derived via both techniques for 54 of the SNe 
in our sample is shown in Figure \ref{f4}.  In general, there is good 
agreement between the two reddening determinations, with no 
evidence for a difference in zero points (the weighted average of the 
difference amounts to 0.01 $\pm$ 0.01).  
The most discrepant points correspond to SNe~1992ag and 1996ai (not plotted).
SN~1992ag, which was discovered and observed in the Cal\'{a}n/Tololo survey,  
occurred near the center of an edge-on spiral host galaxy, and we 
suspect that the B photometry obtained at late epochs was systematically 
affected by the difficulty of subtracting the host galaxy.  A similar
explanation may apply to the severely reddened SN~1996ai for which
we find $E(B-V)_{Template} = 1.97\pm0.08$ and $E(B-V)_{Direct} = 2.43\pm0.19$;
in this case, the $E(B-V)_{Direct}$ estimate is based on a single color
measurement which we suspect may be more uncertain than the formal
photometry errors would suggest.

The zero point of these color excesses can be checked by examining the 
reddenings obtained for the subset of SNe~Ia with E/S0 host galaxies, where 
we expect there to be little or no host galaxy dust\footnote{Although there
is some evidence for a diffuse component of dust in nearby E galaxies that
would be difficult to detect directly (e.g., see \cite{Gou95}), the SNe
with E hosts in our sample occurred at relatively large projected distances
($\geq10$ kpc) from the centers of their host galaxies where the maximum
amount of reddening consistent with observed color gradients is typically
$E(B-V) \leq 0.05$ (\cite{Wis96}).  Thus, it seems reasonable to expect
that SNe in E galaxies should give a reliable determination of the
reddening zero point.}.
For the 21 SNe~Ia in 
our sample with E/S0 hosts, we defined $E(B-V)_{Tail}$ to be the weighted 
mean of the host galaxy color excesses $E(B-V)_{Template}$ and 
$E(B-V)_{Direct}$\footnote{In taking a weighted mean of $E(B-V)_{Direct}$ and
$E(B-V)_{Template}$, we have assumed these to be independent measurements
of the reddening.  Technically speaking, this is not fully the case since 
the tail
photometry used to measure $E(B-V)_{Direct}$ is a subset of the entire
phoometry set which we use in calculating $E(B-V)_{Template}$.  However, for
a typical SN in our sample, the tail data amount to less than half of the
full set of photometry and enter into the template fits with low weight due
to their larger measurement errors.}.  
We find an average value of $E(B-V)_{Tail} = 0.05 \pm 0.03$, which decreases 
to $0.03 \pm 0.03$ if we limit the sample to the eight SNe in E galaxies. 
We do not consider either of these results to be significantly different 
from zero.  Moreover, there are additional arguments for retaining the
Lira zero point:

1. The SNe used to determine the Lira zero point were all relatively
nearby.  We therefore have precise morphological classifications for the
host galaxies.  Being nearby, it's also easier to see signs of dust
directly in the host galaxies.  In addition, we have high signal-to-noise
spectra which allow us to put limits on the amount of interstellar gas
present in the line of sight.  Although Lira used only four SNe for her zero
point, the six shown in Figure~\ref{f1} give the same zero point to 0.01 mag.

2. Of the 21 SNe in our sample with E/S0 host galaxies, 
many are relatively distant.  Some of
the morphological classifications of the hosts could therefore be in error
(e.g., at large distances it's sometimes difficult to distinguish an S0 
galaxy from an Sa).  Hence, we can't be certain that some of these
galaxies don't have significant dust.  Also, very few of the spectra
obtained of these SNe have sufficient signal-to-noise to look for interstellar
absorption lines.

Thus, we adopt the zero point of equation~\ref{eq:eq1} without modification.
In any case, it should be recalled that the precision 
of any single measurement of $E(B-V)_{Tail}$ will be limited by the 
intrinsic dispersion about equation~\ref{eq:eq1} which, as stated earlier,
appears to be $\sim$0.04-0.06~mag.

\subsection{$E(B-V)_{Max}$ and $E(V-I)_{Max}$}

The ``tail'' evolution of the intrinsic $B-V$ color as an indicator of host
galaxy dust reddening offers the advantage of being insensitive to 
the initial decline rate (or luminosity) of the SN, but the obvious 
disadvantage of this method is that it requires the light curves to be 
followed to several magnitudes below maximum light, which in practice is 
often impossible for distant SNe~Ia.  Ideally, we would prefer to be able 
to use the maximum-light colors to determine the host galaxy reddening.  As
is seen in Figure \ref{f1} (and pointed out by \cite{Lir95}), the $B-V$
colors at maximum light of most SNe~Ia show a fairly small dispersion,
although the fastest declining events (such as 1991bg) are clearly 
significantly redder.  In principle, we can calibrate any dependence on
decline rate by using the ``tail'' method to identify a sample of 
unreddened SNe~Ia representing a wide range of initial decline rates.

Columns 2--4 of Table \ref{t2} list the measured decline rate parameter
$\dmm_{obs}$, the color 
excess $E(B-V)_{Gal}$ due to Galactic reddening, and the $E(B-V)_{Tail}$ 
values calculated as per \S~2.1 for the 62 SNe~Ia in our full sample.
We find that a total of 20 events have values of $E(B-V)_{Tail} < 0.06$ ---
i.e., less than or equal to the 
standard deviation of a single reddening measurement.
In the following, we shall assume that 
these SNe~Ia were essentially unreddened by dust in their host galaxies.  
The ``colors'' $B_{max}-V_{max}$ and $V_{max}-I_{max}$ for these 
20 events are plotted versus the decline rate parameter \dmm\ in 
Figure \ref{f5}.  Here we differentiate between those SNe~Ia with light 
curve coverage beginning $< 7$ days after the epoch of $B_{max}$ and those 
for which the first observation was not obtained until $> 7$ days after 
the epoch of
$B_{max}$ since we expect that the color excesses determined for the 
latter SNe will be less precise than those for events observed within a 
week of maximum light.  For decline rates in the range  
$0.9 \leq \dmm \leq 1.6$, the variation in color is 
approximately linear
in both colors.  The fastest-declining events, 1991bg and
1992K, clearly do not follow this trend, presumably due to 
the appearance in 
their spectra of strong absorption features of low-ionization species such 
as Ti~II as a consequence of lower effective temperatures 
(see \cite{Nug_etal95}). 

Over the linear portion of the relations shown in Figure \ref{f5}, we 
calculate the following weighted, least squares fits to the SNe~Ia with 
light curve coverage beginning $< 7$ days after $B_{max}$:

\begin{equation}
(B_{max}-V_{max})_\circ=-0.070(\pm0.012)+0.114(\pm0.037)[\dmm-1.1], 
\sigma=0.030, {\rm n}=14  \label{eq:eq7}
\end{equation}

\begin{equation}
(V_{max}-I_{max})_\circ=-0.323(\pm0.017)+0.250(\pm0.056)[\dmm-1.1], 
\sigma=0.042, {\rm n} = 11 \label{eq:eq8} 
\end{equation}

\noindent
These fits are plotted as solid lines in Figure \ref{f5}. For reference, 
we also show the relations derived by \cite{Phi93}, \cite{Ham_etal96a}, 
the original formulation of MLCS 
(\cite{Rie_etal96a}) and the revised version of \cite{Rie_etal98}.  
Clearly the slopes of the $B_{max}-V_{max}$ relations found by 
\cite{Phi93} and the original MLCS are too steep, whereas both 
the \cite{Ham_etal96a} and the new MLCS provide much better approximations 
to the data.  The small offset between our fit and that of
\cite{Rie_etal98} is due in part to our usage of the \cite{Sch_etal98}
Galactic reddenings, which are on average $\sim0.02$ mag greater than
the \cite{Bur82} values employed by \cite{Rie_etal98}; the larger offset
with respect to the \cite{Ham_etal96a} fits is due to the same effect, and
also to the fact that \cite{Ham_etal96a} did not correct their
data for host galaxy extinction.
Curiously, the \cite{Phi93} relation for $V_{max}-I_{max}$ is a
poor fit to the data, whereas the original MLCS reproduces the 
dependence quite well.

To use the relations given by equations~\ref{eq:eq7} and \ref{eq:eq8} 
to estimate host galaxy reddenings, we procede in a fashion analagous to
that described in \S~2.1.  The observed color excess is defined as

\begin{equation}
E(B-V)_{host}=(B_{max}-V_{max})_{corr}-(B_{max}-V_{max})_\circ, \label{eq:eq9}
\end{equation}

\noindent
where $(B_{max}-V_{max})_{corr}$ is given by

\begin{equation}
(B_{max}-V_{max})_{corr}=(B_{max}-V_{max})_{obs}-E(B-V)_{Gal}-
K_{B_{max}-V_{max}}, \label{eq:eq10}
\end{equation}

In principle, since $\dmm_{obs}$ is a function of the reddening,
equations~\ref{eq:eq9} and \ref{eq:eq10} 
should be employed in an iterative fashion, but the slopes of 
equations~\ref{eq:eq7} and \ref{eq:eq8} are sufficiently shallow that this 
effect can be ignored for all but the most heavily reddened events.  The
observed color excesses are then converted to ``true'' color excesses
via the following approximate relations:

\begin{equation}
E(B-V)_{true} = 0.981 / (1/E(B-V)_{obs} - 0.050) \label{eq:eq11}
\end{equation}

\begin{equation}
E(V-I)_{true} = 0.989 / (1/E(V-I)_{obs} - 0.004) \label{eq:eq12}
\end{equation}

\noindent
Note that these color excesses, which we shall
refer to as $E(B-V)_{Max}$ and $E(V-I)_{Max}$, are valid only for
SNe~Ia with decline rates in the range $0.9 \leq \dmm \leq 1.6$
which also show ``normal'' spectra at
maximum light -- i.e., with strong Si~II~$\lambda$6355 absorption.  In
particular, this method should not be used to calculate reddenings
for SN~1991T-like events which show very different spectra at maximum
and therefore may have dissimilar intrinsic colors (e.g., see 
\cite{Phi_etal92}).

Values of $E(B-V)_{Max}$ and $E(V-I)_{Max}$ for the SNe in our sample
are listed in columns 5--6 of 
Table \ref{t2}, and are plotted versus $E(B-V)_{Tail}$ in Figure \ref{f6}.
Except for SN~1992ag, which we have already discussed, the agreement 
between $E(B-V)_{Max}$ and $E(B-V)_{Tail}$ is generally excellent.  
The relationship between $E(V-I)_{Max}$ and $E(B-V)_{Tail}$, while clearly 
significant, is not as impressive.  Perhaps this is not surprising in 
view of the larger dispersion observed in the $E(V-I)_{Max}$ versus 
\dmm~relation for relatively unreddened SNe~Ia (see Figure \ref{f5} and
equation \ref{eq:eq8}) which we 
suspect arises from two main causes:  the generally poorer sampling of 
the $I$-band light curves for many of the Cal\'{a}n/Tololo SNe, and the more 
complicated nature of the shapes of the $I$-band light curves which makes 
predicting $I_{max}$ from template fitting challenging.  Nevertheless, we 
consider the $V_{max}-I_{max}$ color to be a useful alternative method 
of identifying significantly reddened SNe~Ia\footnote{Strictly speaking,
color excesses estimated via equations~\ref{eq:eq7} and \ref{eq:eq8} are
not fully independent of the $E(B-V)_{Tail}$ measurements for the same
SNe since a change in the zero point or slope of the Lira relation
(equation~\ref{eq:eq1}) could change somewhat the exact sample of unreddened
events used to derive equations~\ref{eq:eq7} and \ref{eq:eq8}.  Also, an
uncertainty in $V_{max}$ will produce a correlated error in $E(B-V)_{Max}$ and
$E(V-I)_{Max}$ since both use the same measurement.}.

In column 7 of Table \ref{t2} we list the average color excess $E(B-V)_{Avg}$ 
which we take to be the weighted mean of $E(B-V)_{Tail}$, $E(B-V)_{Max}$, 
and $0.8E(V-I)_{Max}$.  The multiplicative factor of 0.8 for
$E(V-I)_{Max}$ corresponds to an assumption of a standard Galactic 
interstellar reddening curve (\cite{Dea_etal78}; see also \S~3).  
In addition, to handle negative reddening values we have applied the 
``Bayesian filter'' 
used by \cite{Rie_etal98} which assumes a one-sided Gaussian {\it a priori} 
distribution of $A_B$ values with a maximum at zero and 
$\sigma = 0.3$ mag.  For the subsample of 19 SNe~Ia in 
E and S0 host galaxies with $0.9 \leq \dmm \leq 1.6$, we find

\begin{displaymath}
E(B-V)_{Avg}=0.02\pm0.03, \sigma=0.05, {\rm n}=21,
\end{displaymath}

\noindent
which implies that the zero points of equations~\ref{eq:eq1}, \ref{eq:eq7} 
and \ref{eq:eq8} are reasonable, and that the typical precision of a
measurement of $E(B-V)_{Avg}$ for any single SN~Ia is $\sim$0.05~mag.  
As expected, we find that the most reddened 
SNe were observed in spiral galaxies: 82\% of the SNe with host galaxy 
reddenings greater than 0.05~mag occurred in spirals, whereas only 29\% of 
the SNe with $E(B-V)_{Avg} \leq$ 0.05~mag had spiral galaxy hosts.

Figure \ref{f7} shows the total reddening, $E(B-V)_{Avg}$, derived for the 
Cal\'{a}n/Tololo SNe~Ia plotted as a function of host galaxy redshift.  
From the Hubble diagram fits given in \S~3 and the definition of the 
observed magnitude of a supernova as $B_{obs} = B_o + A_B$, it can easily 
be shown that the maximum allowed color excess for a SN as a function of 
redshift in a magnitude-limited search is 

\begin{equation}
E(B-V) \simeq (B_{limit} - 5log(cz) + 3.57) / 4.1. \label{eq:eq13}
\end{equation}

\noindent
The dashed line in Figure \ref{f7} shows equation~\ref{eq:eq13} plotted for 
the limiting magnitude of the Cal\'{a}n/Tololo search, $B_{limit} \sim 19$ 
(\cite{Ham_etal93}).  This curve
provides a good fit to the upper envelope of the observed reddenings, 
particularly at higher redshifts.  The fact that SNe in a magnitude-limited 
search will have preferentially smaller dust reddenings as a function of 
redshift will naturally lead to a greater percentage of SNe being 
discovered in E/S0 galaxies at higher redshifts, exactly as is observed 
in the Cal\'{a}n/Tololo sample (\cite{Ham_etal96b}).  
These results and predictions 
provide confidence in the validity of our techniques for measuring the 
reddenings of SNe~Ia.

\section{The Reddening-Free Decline Rate versus Luminosity Relation}

Once the reddening of a SN~Ia is known, correcting 
the observed maximum-light magnitudes for this extinction is straight-forward.
The general formula for a given filter is

\begin{displaymath}
m_\circ = (m_{obs} - A_m(Gal)) - K_m - A_m(host),
\end{displaymath}

\noindent
where $m_{obs}$ is the observed maximum-light magnitude, $A_m(Gal)$ is the
Galactic component of extinction, $K_m$ is the K~correction (which, as noted
in \S~2.1, is a function of both redshift and reddening), and $A_m(host)$ is
the host galaxy component of extinction.  From the data shown in 
Figure \ref{f2} and similar calculations for the $V$ and $I$ bands, 
the dependence of the extinction on 
color excess at maximum light can be approximated as

\begin{equation}
A_B = [4.16 - 0.06 E(B-V)_{true}] E(B-V)_{true}, \label{eq:eq14}
\end{equation}

\begin{equation}
A_V = [3.14 - 0.02 E(B-V)_{true}] E(B-V)_{true}, \label{eq:eq15}
\end{equation}

\noindent
and

\begin{equation}
A_V = [1.82 + 0.01 E(B-V)_{true}] E(B-V)_{true}. \label{eq:eq16}
\end{equation}

\noindent
Equations~\ref{eq:eq14}-\ref{eq:eq16} are valid for calculating either
the Galactic or host galaxy components of the extinction.

The left half of Figure \ref{f8} shows the absolute magnitudes corrected 
only for Galactic reddening for the full sample of 29~Cal\'{a}n/Tololo and 
12~CfA SNe with $z \geq 0.01$ plotted as a function of the observed 
decline rate parameter $\dmm_{obs}$.  
The distance to each SN was calculated from the redshift 
of the host galaxy (in the CMB frame) and an assumed value of the Hubble 
constant of \ho = 65 \kms Mpc$^{-1}$.  Following \cite{Ham_etal96a}, a
peculiar velocity term of 600 \kms\ has been included in the error bars of
the absolute magnitudes.

In the right half of Figure \ref{f8}, 
this plot is repeated for the subsample of 18 of these 41 SNe for 
which we find
insignificant host galaxy reddening ($E(B-V)_{Avg} < 0.05$).  This 
diagram reveals more clearly than ever the true nature of the peak 
luminosity-decline rate relations for SNe~Ia.  Elimination of the events 
with significant host galaxy reddening not only decreases the scatter in 
the relations, but also shows these to be nonlinear in the $B$ and $V$ bands,
and probably also in $I$.  
(Note that \cite{Rie_etal98} also found this dependence to be nonlinear
and used a quadratic term in the revised MLCS.)
In the right half of Figure \ref{f8}, we plot 
for comparison as dashed lines the linear fits given 
by \cite{Ham_etal96a} for the 26 Cal\'{a}n/Tololo SNe with 
$B_{max}-V_{max} < 0.2$.  These show best agreement in $I$ where reddening 
effects are smaller, but are progressively worse approximations to the 
actual peak luminosity-decline rate relations in the $V$ and $B$ bands due 
to the combined effects of uncorrected host galaxy extinction and the 
curvature which is apparent in the true relations.

For this ``low host galaxy reddening'' subsample of the Cal\'{a}n/Tololo 
and CfA SNe with decline rates in the range $0.85 < \dmm < 1.70$, we 
calculated a least square fit to a quadratic peak luminosity-decline rate 
relation for the $V$ band data.  This fit was then combined with 
equations~\ref{eq:eq7} and \ref{eq:eq8} to give corresponding relations
for the $B$ and $I$ bands.
Table \ref{t3} lists the resulting 
parameters for these fits expressed in terms of the magnitude difference,
$\Delta M_{max}$, with respect to the maximum light magnitude of a SN with
$\dmm = 1.1$.  
Note that the low values of 
$\chi_\nu^2$ are probably due to an overestimate of the errors 
by \cite{Ham_etal96a} which we have adopted verbatim.
The impressively 
low dispersions (0.09-0.13~mag) of these fits strongly support 
the validity of the methods 
described in this paper for estimating the host galaxy dust reddening.  
For reference, \cite{Ham_etal96a}, who employed a color cut of 
$B_{max}-V_{max} < 0.2$ and assumed linear 
relations for \dmm\ versus $M_{max}$, obtained 
dispersions of 0.17, 0.14, and 0.13~mag for the fits in $B$, $V$, and $I$, 
respectively.  The wavelength dependence of these values is due 
to a combination of uncorrected host galaxy extinction and the non-linear 
shape of the \dmm\ versus $M_{max}$ relation in $B$ and $V$; when 
these effects are taken into account the dispersion is uniformly low in 
$BVI$.

From the average values of the host galaxy reddening given in Table \ref{t2}
and the decline 
rate versus luminosity relations given in Table \ref{t3}, we have reexamined 
the corrected Hubble diagrams in $BVI$ for the 26 
Cal\'{a}n/Tololo with $B_{max}-V_{max} < 0.2$, which Hamuy et al. (1996a, 1996b)
referred to as their ``low-extinction'' sample.  To provide an absolute 
calibration for these diagrams and, hence, an estimate of the Hubble 
constant, we use the same four SNe (1937C, 1972E, 1981B, and 1990N)
with Cepheid distances employed by \cite{Ham_etal96b}.  Three different 
cases were
considered: 1) Correcting for Galactic reddening only, 2) correcting for
Galactic reddening and the \dmm\ versus $M_{max}$ relation, and 3) correcting
for Galactic reddening, host galaxy reddening, and the \dmm\ versus $M_{max}$ 
relation.

Averaging the results for $BVI$, we find the Hubble relations
illustrated in Figure \ref{f9}.  (Note that the errors given in this 
figure for
\ho~are internal only, and do not include the external error due to 
uncertainties in the zero point of the Cepheid period-luminosity 
calibration, which we estimate to be an additional 3--4 \kms\ Mpc$^{-1}$.)
Not surprisingly, for the first two cases we find Hubble constants and 
dispersions which are similar to those given by \cite{Ham_etal96b} for 
comparable corrections.  Note that case 1 is essentially the method employed 
by \cite{San_etal96}, who ignored the existence of the \dmm\ versus $M_{max}$ 
relation; the lower Hubble constant results from the fact that the average 
decline rate of the 4 calibrators with Cepheid distances (\dmm\ = 0.97)
is significantly lower than the mean value for the 26 Cal\'{a}n/Tololo SNe
(\dmm\ = 1.24), meaning that the 
calibrators are on average $\sim0.2$ mag more luminous than the 
Cal\'{a}n/Tololo SNe.  Correcting for the \dmm\ versus $M_{max}$ relation
(case 2) eliminates this bias and significantly decreases the dispersion
in the Hubble diagram.  Applying the host galaxy reddening corrections
(case 3) decreases the dispersion still further, illustrating the merit of
our approach to calculating these corrections, although we find essentially 
the same value of \ho~as when this correction is ignored (case 2).
This is ascribed to the coincidence that the median reddening 
($E(B-V)_{host} = 0.05$~mag) of the 26 distant SNe 
happens to be virtually identical 
to the median reddening ($E(B-V)_{host} = 0.06$ mag)
for the four calibrating events.

From fits to the fully corrected (i.e., for Galactic + host galaxy
reddening, and the \dmm\ versus $M_{max}$ relations) magnitudes of our
complete sample of 28 Cal\'{a}n/Tololo and 12 Cfa SNe with $z \geq 0.01$
and $\dmm < 1.70$, we find the following relations for the Hubble constant
in $BVI$:

\begin{eqnarray}
{\rm log}\ho(B)=0.2\{M^B_{max}-0.786(\pm0.398)[\dmm-1.1]- \nonumber \\
0.633(\pm0.742)[\dmm-1.1]^2+28.671(\pm0.043)\}, \label{eq:eq17}
\end{eqnarray}

\begin{eqnarray}
{\rm log}\ho(V)=0.2\{M^V_{max}-0.672(\pm0.396)[\dmm-1.1]- \nonumber \\
0.633(\pm0.742)[\dmm-1.1]^2+28.615(\pm0.037)\}, \label{eq:eq18}
\end{eqnarray}

\begin{eqnarray}
{\rm log}\ho(I)=0.2\{M^I_{max}-0.422(\pm0.400)[\dmm-1.1]- \nonumber \\
0.633(\pm0.742)[\dmm-1.1]^2+28.236(\pm0.035)\}. \label{eq:eq19}
\end{eqnarray}

\noindent
Combining equations \ref{eq:eq17}-\ref{eq:eq19} with the
corrected absolute magnitudes of the six well-observed SNe~Ia (1937C, 1972E,
1981B, 1989B, 1990N, 1998bu --- see \cite{Sah_etal99} and \cite{Sun_etal99} 
for details and
references) with HST Cepheid distances gives a value of the Hubble constant of 
63.3 $\pm$ 2.2(internal) $\pm$ 3.5(external) \kms\ Mpc$^{-1}$. 
Until more Cepheid distances become available for nearby, well-observed 
events, this value represents our best estimate for the Hubble constant.

Finally, we briefly address the question of whether the extinction law of the 
host galaxy dust affecting SNe~Ia is consistent with that typically observed 
for dust in the disk of our own Galaxy.  Using their original formulation of
MLCS, \cite{Rie_etal96b} concluded that this {\em was} the case, but since
this same version of MLCS incorrectly predicted a strong relation between light 
curve shape and the $B-V$ color at maximum light, it is worth reexamining
the issue.  We do this by plotting the absolute
magnitudes of the SNe in our sample versus the $B_{max}-V_{max}$ color.
However, since both of these quantities are a function of the decline rate we
must first correct them to a ``standard'' decline rate (we choose \dmm = 1.1)
using the coefficients of the decline-rate-dependent portions of the
relations given in Table \ref{t3} and equation~\ref{eq:eq7}.  
Specifically, we calculate,

\begin{equation}
M(B)_{\it 1.1}=M(B)_{max}+0.786[\dmm-1.1]+0.633[\dmm-1.1]^2, \label{eq:eq20}
\end{equation}

\begin{equation}
M(V)_{\it 1.1}=M(V)_{max}+0.672[\dmm-1.1]+0.633[\dmm-1.1]^2, \label{eq:eq21}
\end{equation}

\begin{equation}
M(I)_{\it 1.1}=M(I)_{max}+0.422[\dmm-1.1]+0.633[\dmm-1.1]^2, \label{eq:eq22}
\end{equation}

\noindent
and

\begin{equation}
(B_{max}-V_{max})_{\it 1.1}=(B_{max}-V_{max})-0.114[\dmm-1.1]. \label{eq:eq23}
\end{equation}

\noindent
These quantities are plotted in Figure \ref{f10} for the subsample of 
Cal\'{a}n/Tololo and CFA SNe~Ia with z $\geq 0.01$ for which distances were
calculated assuming \ho=63.3, and also for the SNe~Ia with Cepheid-based host 
galaxy distances.  In both cases, the samples were restricted to those events
with light curve coverage beginning $\leq 5$ days after $B_{max}$ and for
which we have derived non-zero values of $E(B-V)_{Avg}$.  Plotted in
Figure \ref{f10} for reference are the canonical Galactic reddening vectors 
with slopes of 4.1 in $B$, 3.1 in $V$, and 1.8 in $I$.  A linear least squares
fit to the data gives slopes of $3.5\pm0.4$, $2.6\pm0.4$, and $1.2\pm0.4$,
which agree to within $\sim1-1.5~\sigma$ with the Galactic values, and which are
consistent with typical values of $R$ observed in nearby galaxies 
(e.g., \cite{Hod82,Bou_etal85,Iye85,Bro88}).  Hence, we conclude that the
host galaxy dust is similar in its properties, on average, to the dust in
our own Galaxy. 

\section{Conclusions}

In this paper we have presented methods for estimating the host galaxy 
dust extinction of SNe~Ia at late times (30 $\leq t_V \leq$ 90), when the 
$B-V$ evolution is remarkably similar for all events, and at maximum light 
where the colors $B_{max}-V_{max}$ and $V_{max}-I_{max}$ for normal-spectra
SNe~Ia display 
only a mild dependence on light curve shape.  From this, we are able to 
deduce the reddening-free relations in $BVI$ for the dependence 
of absolute magnitude on the decline rate parameter \dmm.  Fits to these 
data yield remarkably low dispersions (0.09-0.13~mag) in all three colors, 
and lead to a revised value of the Hubble constant of 
\ho = 63.3 $\pm$ 2.2(internal) $\pm$ 3.5(external) \kms\ Mpc$^{-1}$.

\acknowledgments

We thank Adam Riess for his valuable comments on 
this paper, and for providing the new MLCS fit in Figure \ref{f5}.
J.M. acknowledges support from ``C\'{a}tedra Presidencial
de Ciencias 1995'' and FONDECYT grant No. 1980172.

{}

\newpage 

% FIGURE CAPTIONS

\figcaption[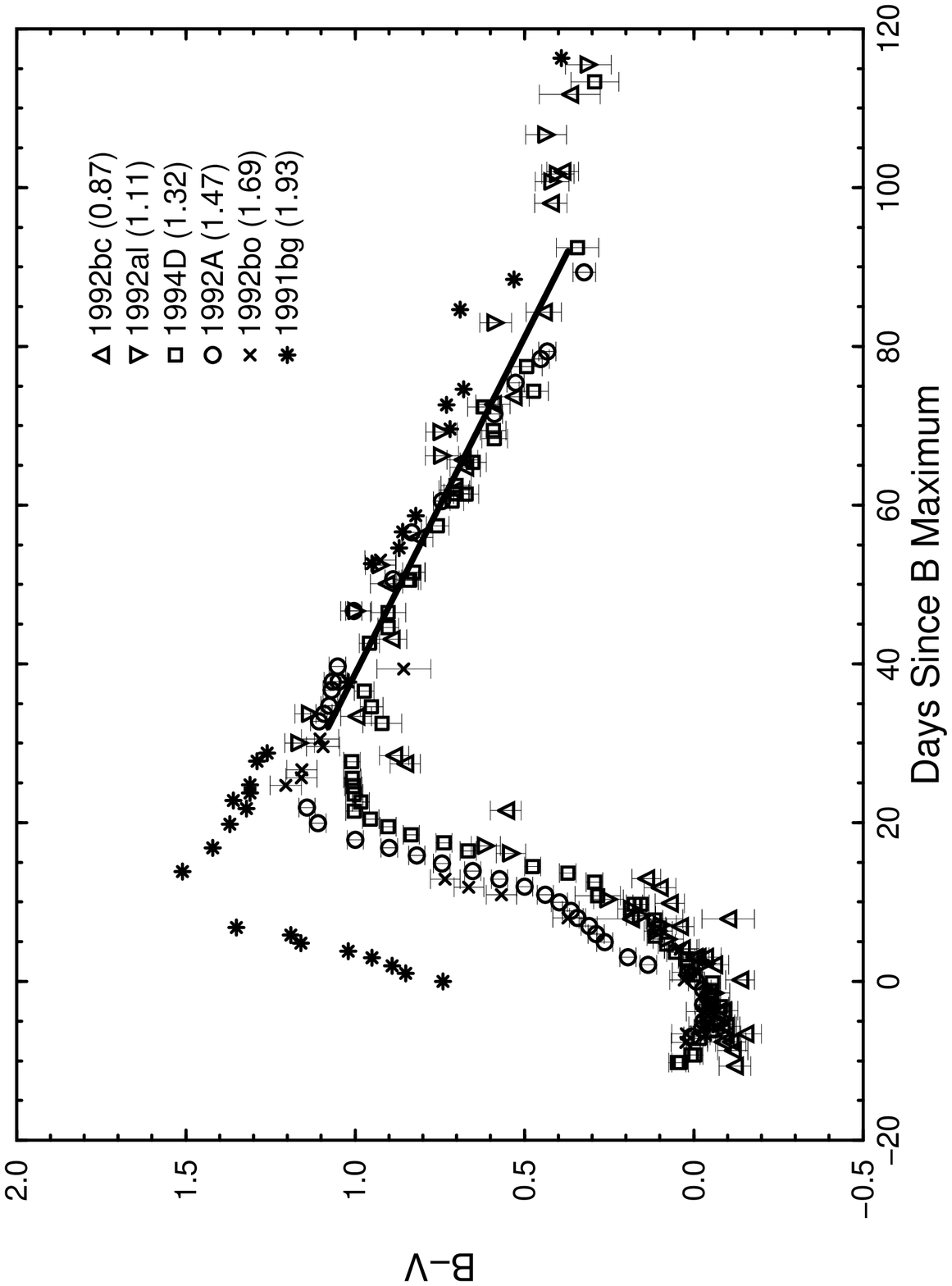]{$B-V$ color evolution for six SNe~Ia which likely 
suffered little or no reddening from dust in their host galaxies.  These 
six events, whose \dmm\ parameters are indicated in parentheses, 
cover a wide range of initial decline rates and peak luminosities.  The 
solid line corresponds to the \cite{Lir95} fit (equation~\ref{eq:eq1}) 
to the color evolution 
during the phase interval 30 $\leq t_V \leq$ 90.  The epoch of $B_{max}$
is assumed to occur 2 days before the epoch of $V_{max}$ (\cite{Lei88}).
\label{f1}}

\figcaption[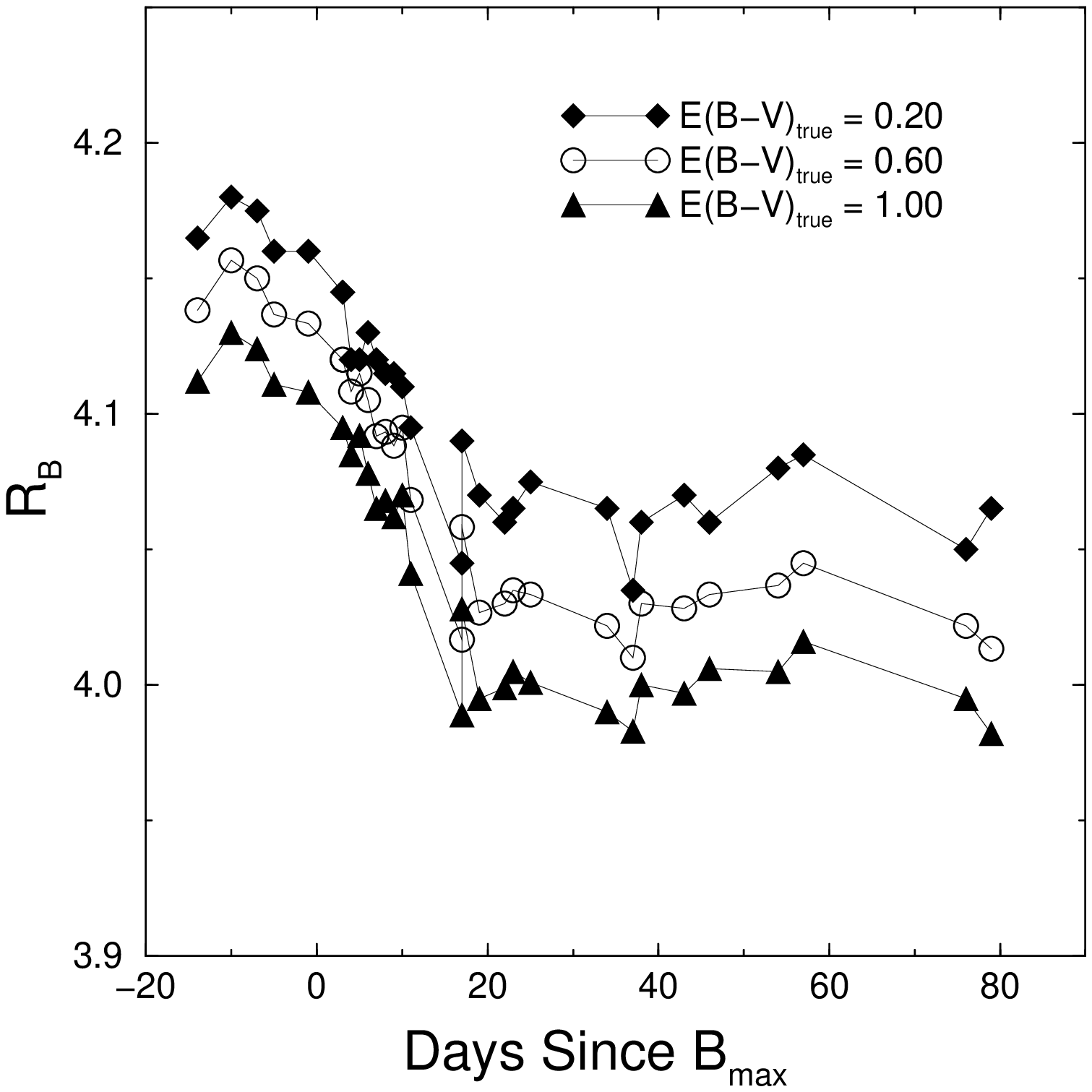]{Variation of $R_B=A_B/E(B-V)_{true}$ as a 
function of phase for different values of $E(B-V)_{true}$. Note that
$A_B$ is the observed extinction in $B$, which is a function of the
light curve phase.\label{f2}}

\figcaption[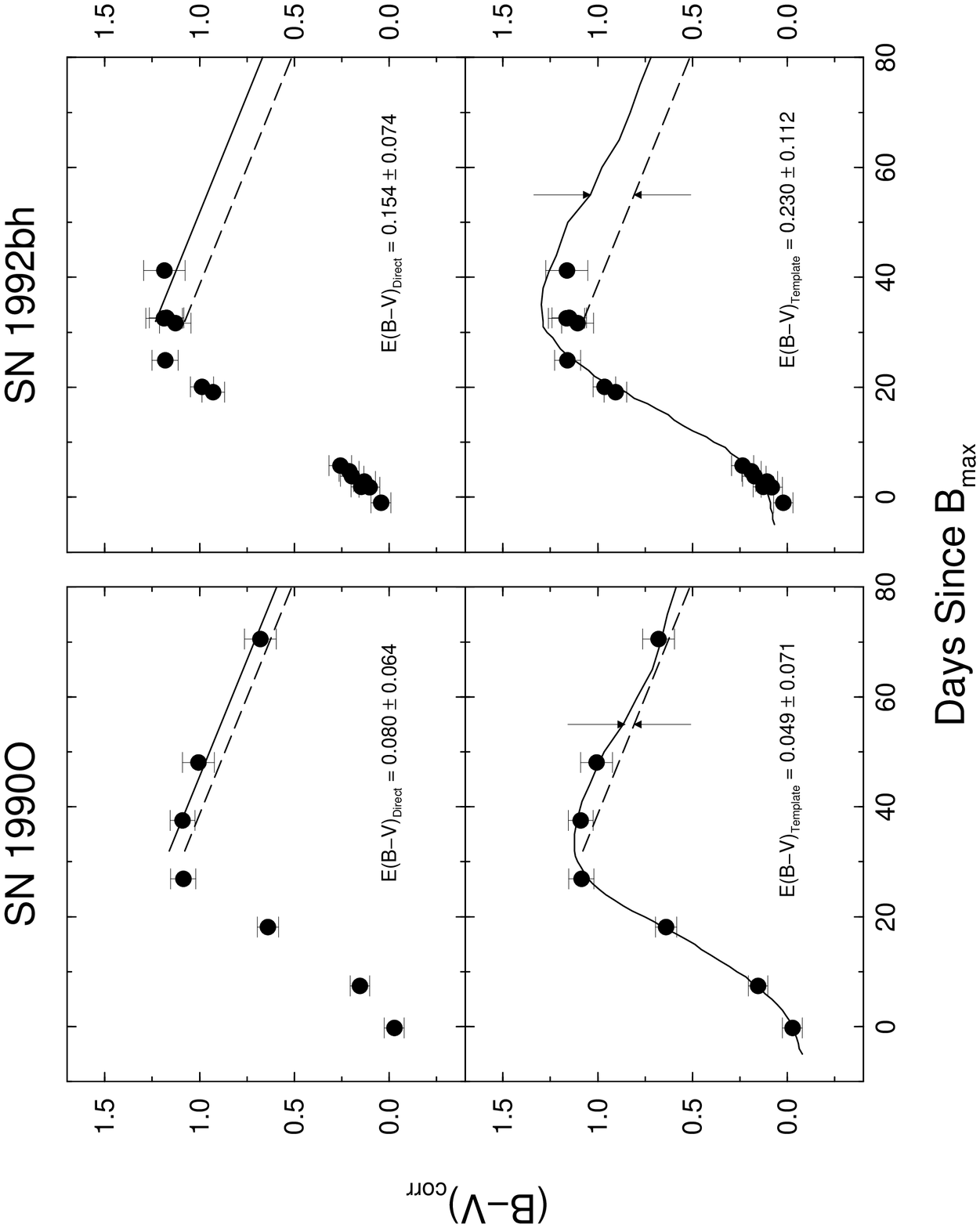]{Illustration of the the ``Direct'' (upper panels)
and ``Template'' (lower panels) techniques for estimating the host 
galaxy reddening of SNe~Ia from the $B-V$ evolution during the
phase interval 30 $\leq t_V \leq$ 90.  The SN photometry
is from \cite{Ham_etal96c} and has been corrected for both Galactic 
reddening and the K~correction.  The dashed line in each graph corresponds
to the \cite{Lir95} fit (equation~\ref{eq:eq1}) to the color 
evolution of unreddened SNe~Ia.
The solid lines in the upper panels are the Lira relation shifted to fit
the photometry of the SNe in the interval 30 $\leq t_V \leq$ 90;
$E(B-V)_{Direct}$ is then defined as the difference between the solid and 
dashed lines.  In the lower panels, the solid lines represent the template
fits derived by \cite{Ham_etal96c} to the corrected photometry.  The
value of $E(B-V)_{Template}$ is defined as the
difference between the template fits and the unreddened Lira relation
(the dashed lines) at $t_V$ = 53 days ($t_B \simeq$ 55 days). \label{f3}}

\figcaption[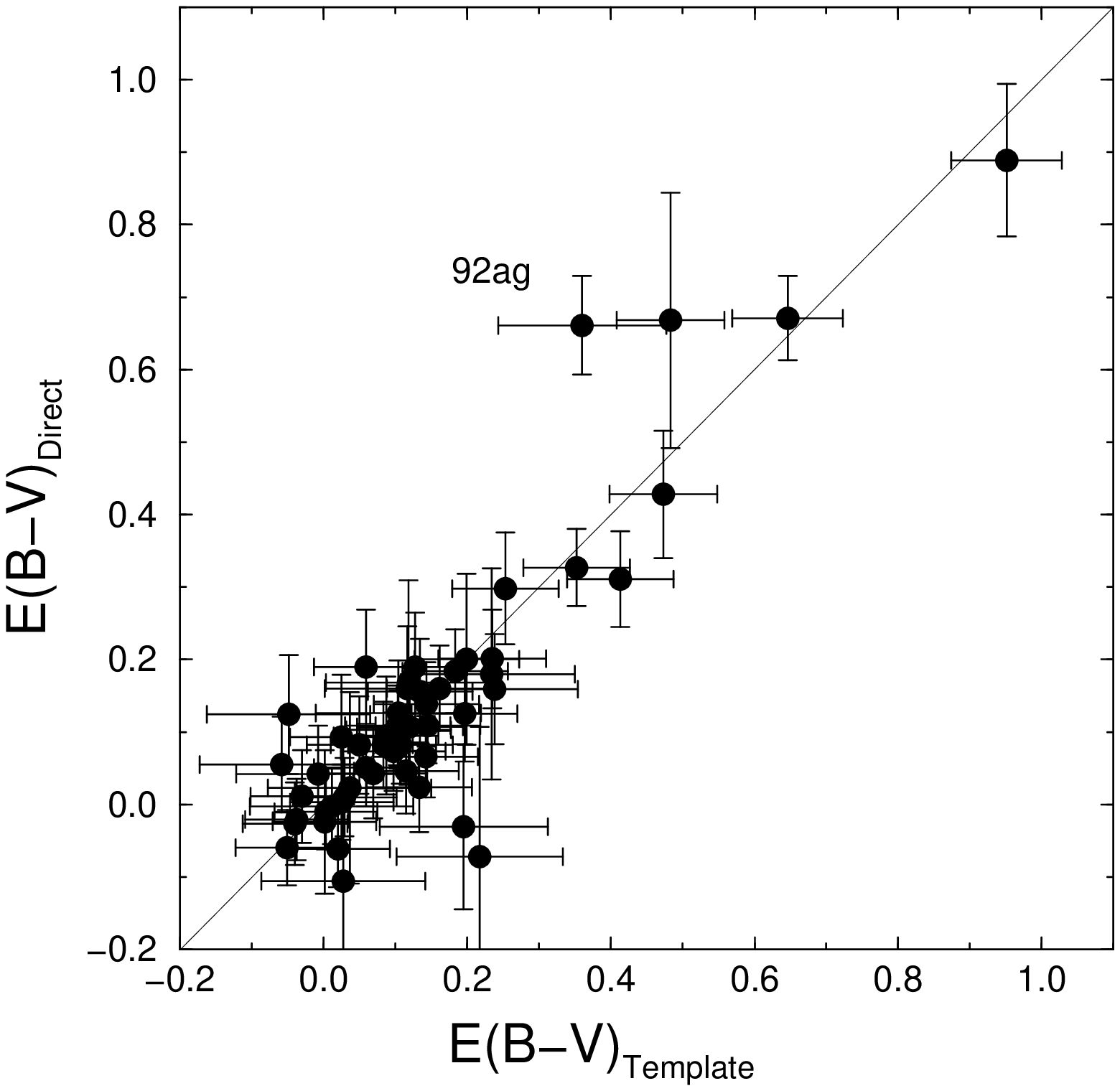]{Comparison of color excesses calculated via the
``Direct'' and ``Template'' methods from the $B-V$ evolution during the 
phase interval 30 $\leq t_V \leq$ 90. \label{f4}}

\figcaption[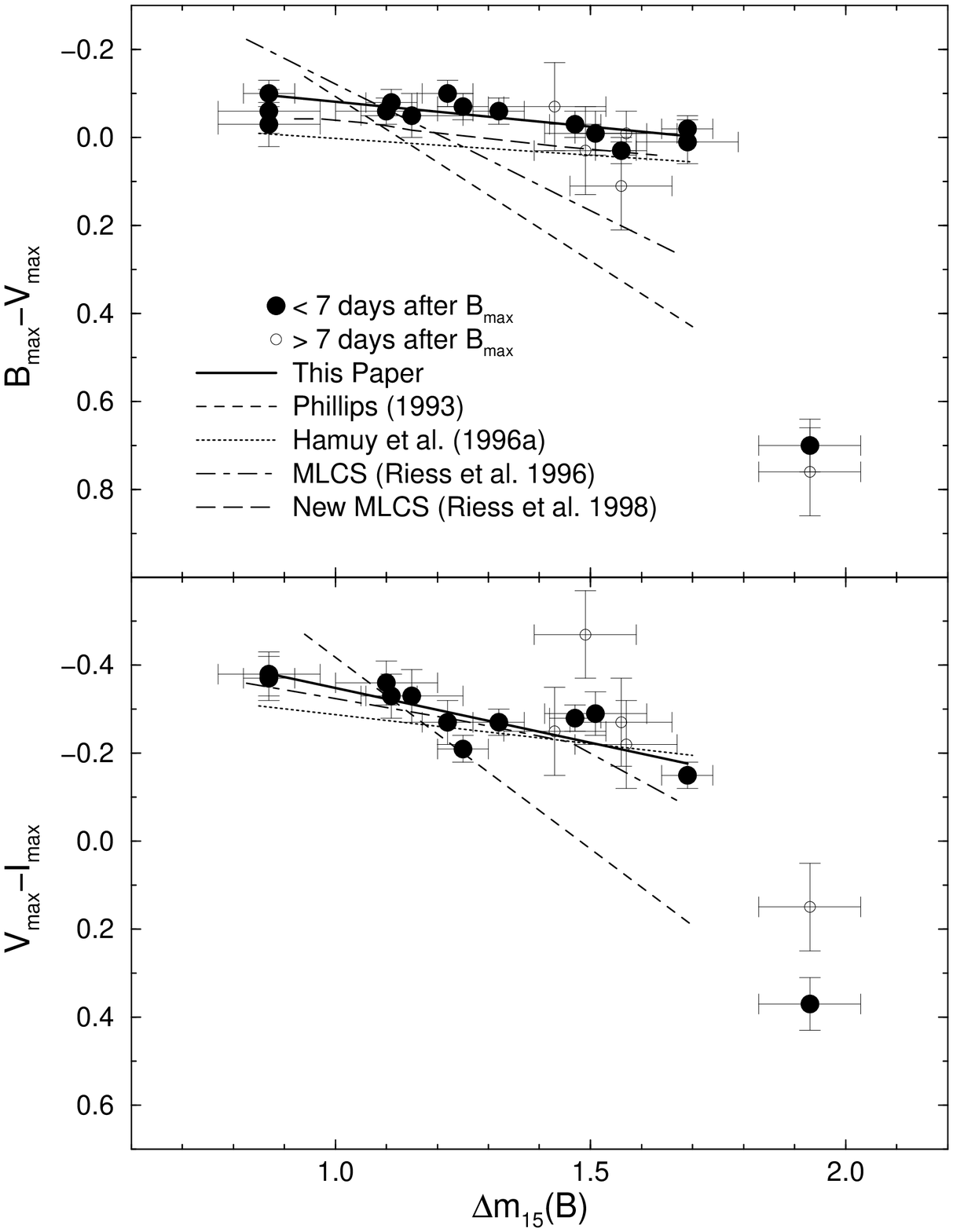]{(Above) The $B_{max}-V_{max}$ ``color'' 
is plotted 
versus the decline rate parameter \dmm\ for 20 events with 
$E(B-V)_{Tail} < 0.06$.  These SNe are likely 
to have little or no host galaxy reddening.  (Below) The $V_{max}-I_{max}$ 
``color''is plotted versus the decline rate parameter \dmm\ 
for the same subsample of essentially unreddened SNe~Ia. \label{f5}}

\figcaption[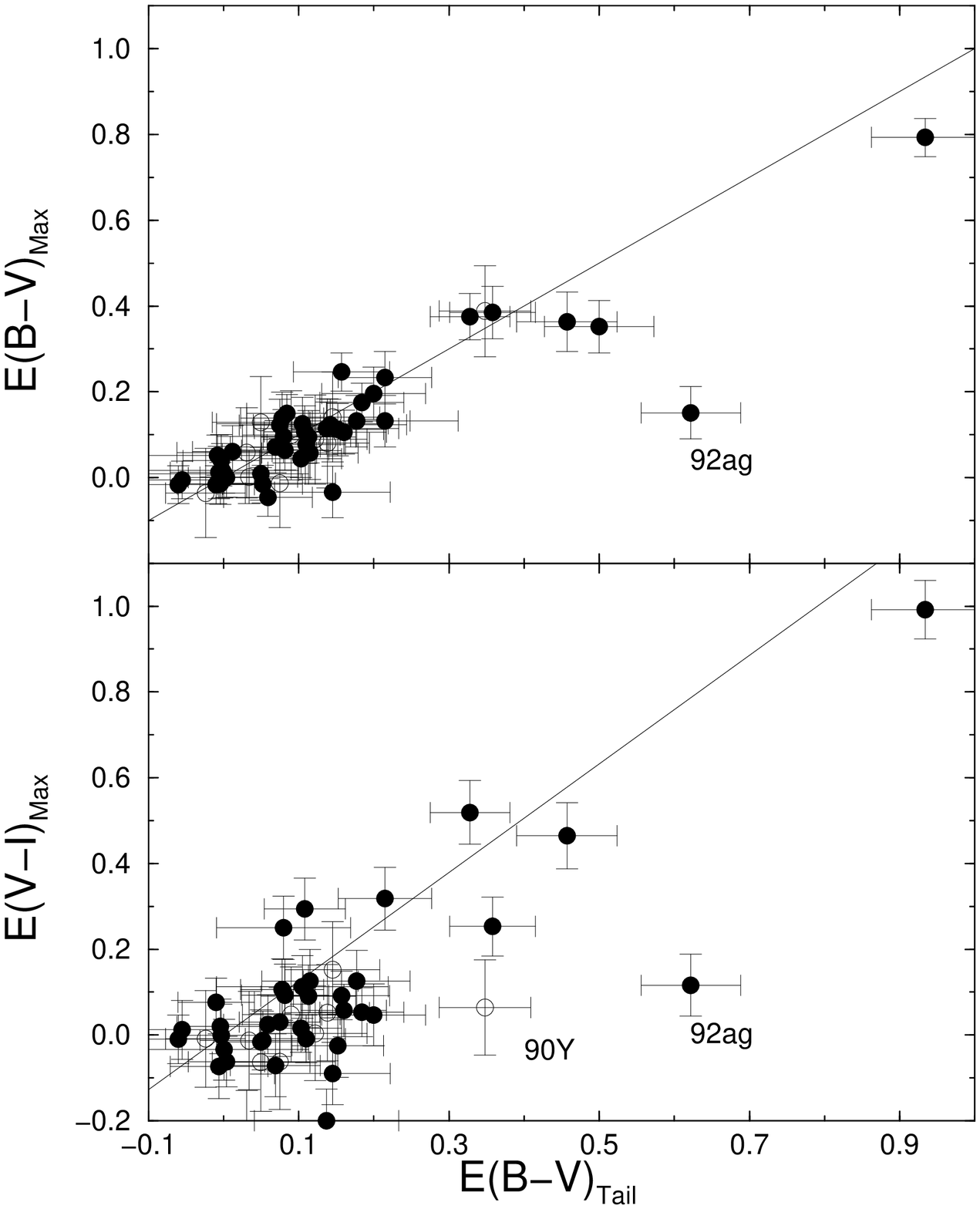]{Comparison of color excesses calculated from the
$B_{max}-V_{max}$ and $V_{max}-I_{max}$ colors versus
$E(B-V)_{Tail}$. The solid line plotted in the lower panel assumes
$E(V-I)_{Max} = 1.25E(B-V)_{Max}$, typical of dust in our own
Galaxy.  Open symbols correspond to SNe for which the first observation was
not obtained until $> 7$ days after the epoch of $B_{max}$. \label{f6}}

\figcaption[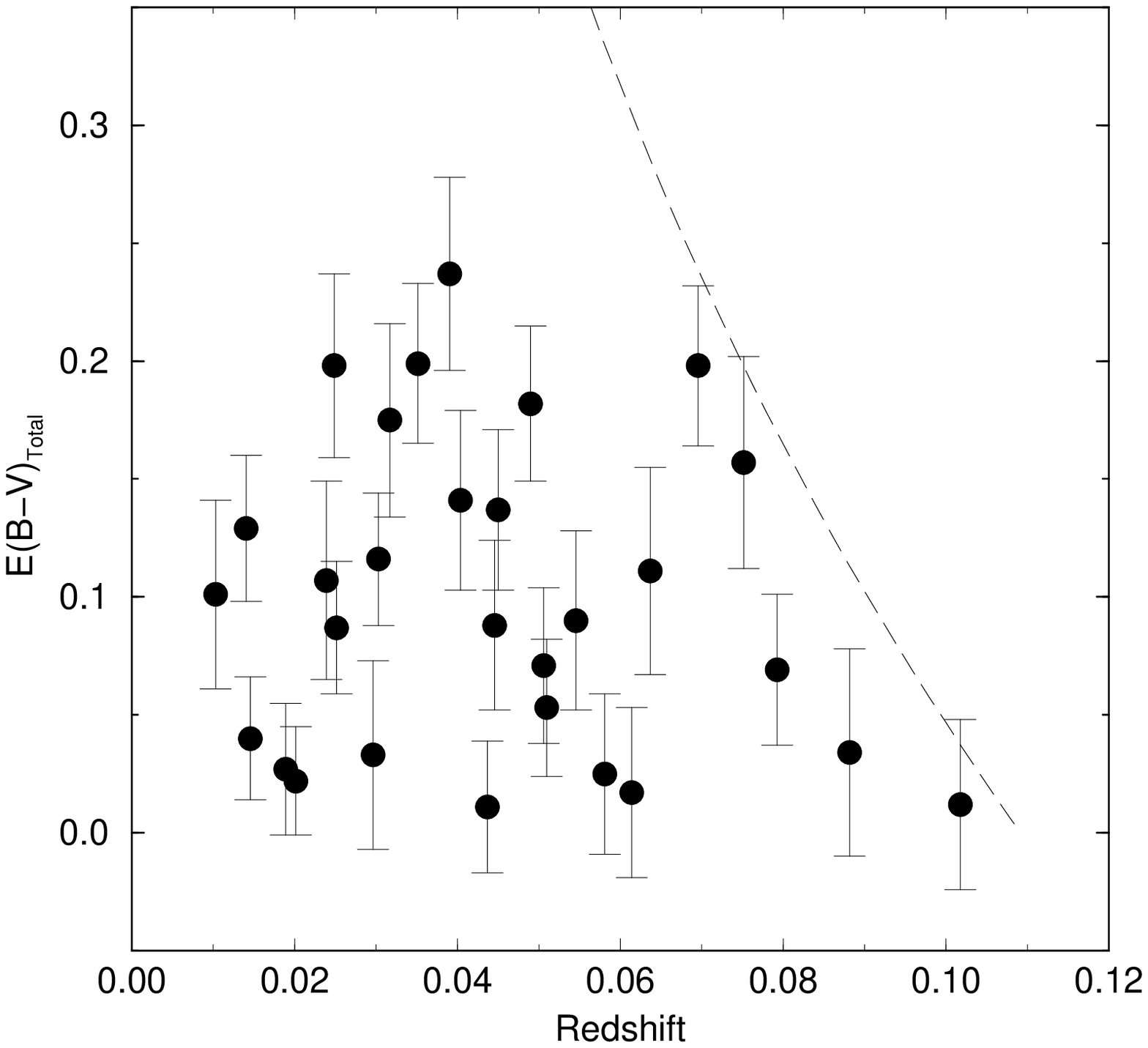]{The total dust reddening (i.e., the sum of the 
Galactic and host galaxy contributions) plotted as a function of redshift
for the 29 SNe~Ia in the Cal\'{a}n/Tololo sample.  The dashed line shows
the upper limit to the allowed color excess for the limiting magnitude of
this search, $B_{Limit} = 19$. \label{f7}} 

\figcaption[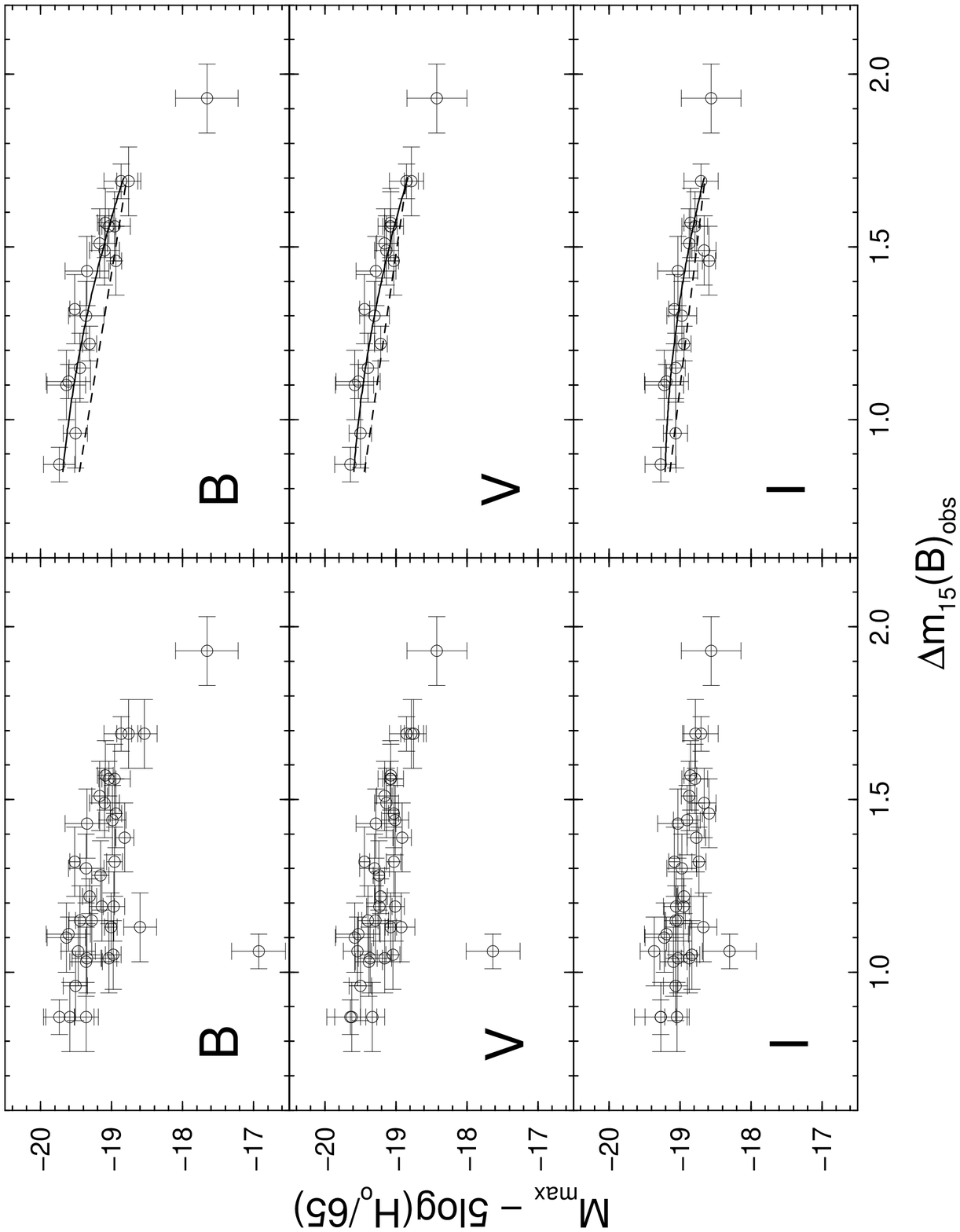]{(Left) The absolute $BVI$ magnitudes 
corrected only for Galactic reddening
plotted versus the decline rate parameter \dmm\ for the full sample of 
41 Cal\'{a}n/Tololo and CfA SNe~Ia with z $\geq 0.01$. (Right) The same 
diagram after elimination of the 23 SNe in the sample for which we find 
significant host galaxy reddening.  The dashed lines correspond to 
the linear fits given by \cite{Ham_etal96a}. \label{f8}}

\figcaption[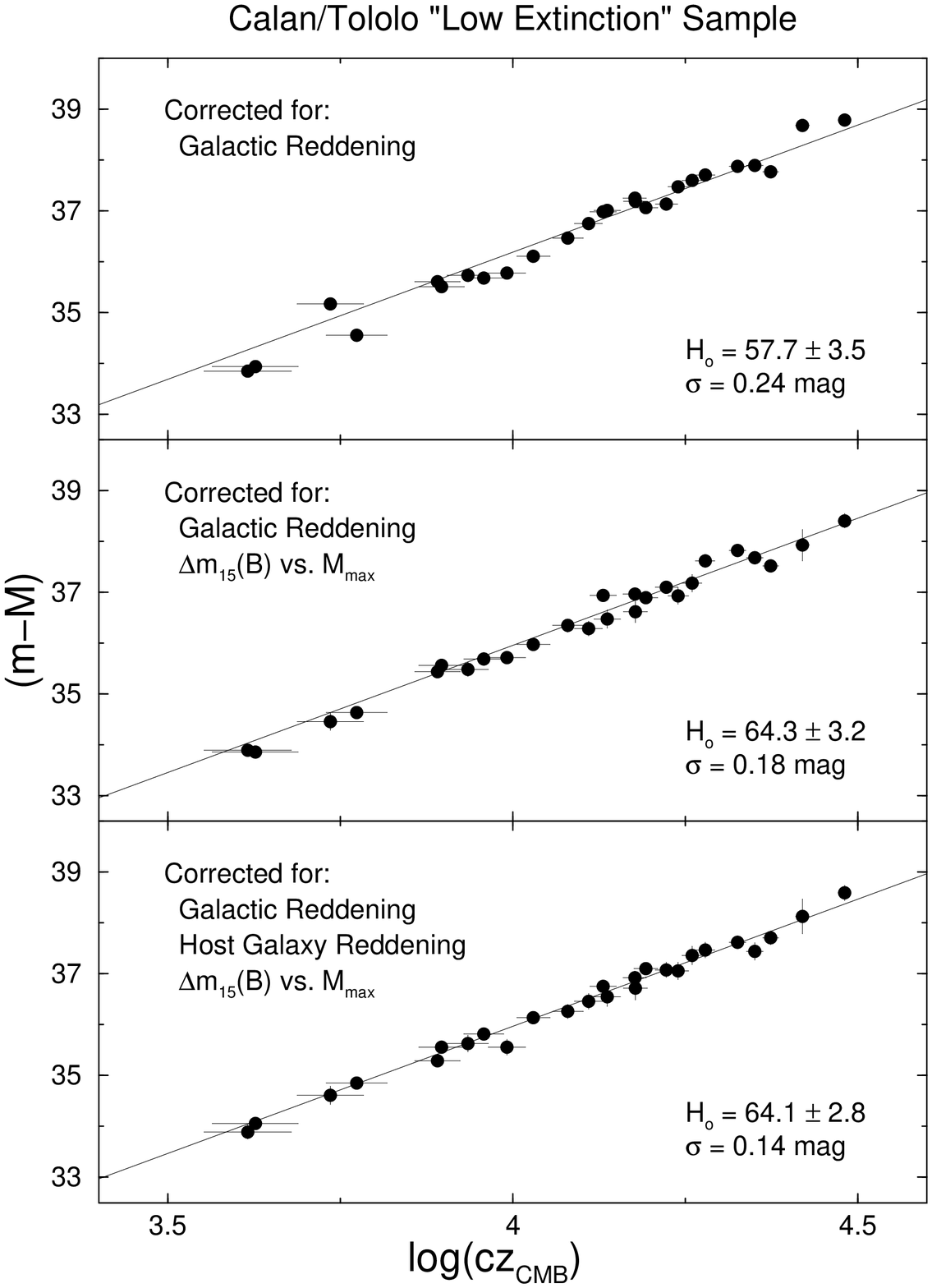]{Hubble diagrams for the 26 SNe~Ia in the 
Cal\'{a}n/Tololo ``low-extinction'' ($B_{max}-V_{max} < 0.2$) sample.
The data are shown corrected for Galactic reddening only (upper panel), 
Galactic reddening and the \dmm\ versus $M_{max}$ relation (middle panel), 
and Galactic reddening, host galaxy reddening, and the \dmm\ versus $M_{max}$
relation (lower panel).  The Hubble constants and dispersions derived
for each case are indicated. \label{f9}}

\figcaption[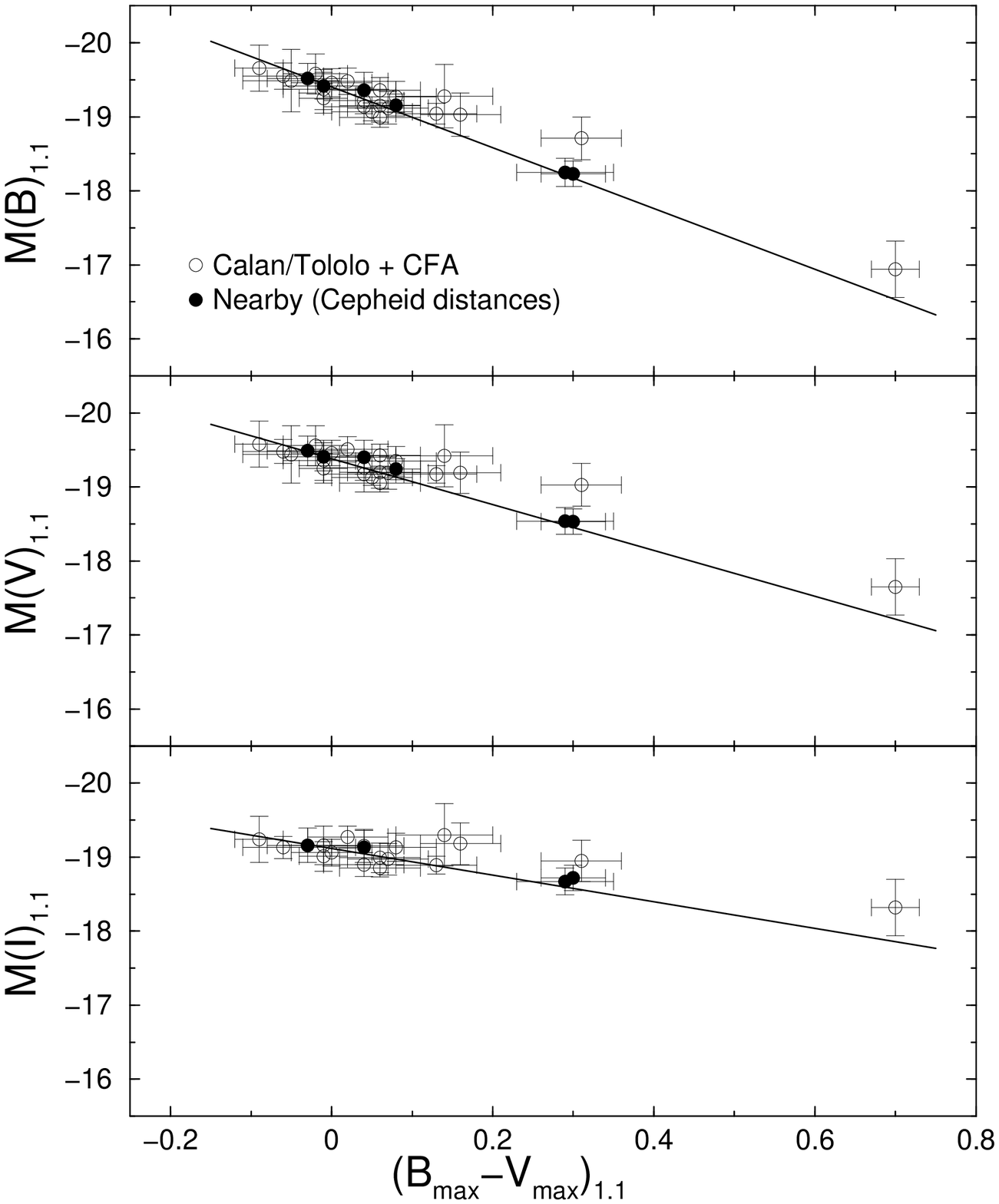]{SN~Ia Absolute magnitudes plotted versus
$B_{max}-V_{max}$ for two samples consisting of the Cal\'{a}n/Tololo +
CFA SNe~Ia with z $\geq 0.01$ and nearby SNe~Ia for which distances have
been determined via Cepheids. Distances to the former set of objects were 
calculated assuming \ho=63.3.  The decline rate dependence of both the
absolute magnitudes and the $B_{max}-V_{max}$ color have been removed via
equations~\ref{eq:eq20}-\ref{eq:eq23}. For both samples, only those SNe
with light curve coverage beginning $\leq5$ days after the epoch of
$B_{max}$ were considered.  The canonical Galactic reddening
vectors are illustrated as solid lines. \label{f10}}

\newpage
\singlespace

\begin{deluxetable}{ll}
\tablenum{1}
\tablewidth{0pt}
\tablecaption{Photometry References for Nearby SNe Ia\label{t1}}
\tablehead{
\colhead{SN} &
\colhead{References}
}
\startdata
1937C  &  Pierce \& Jacoby (1995) \nl
1972E  &  Ardeberg \& de Grood (1973); Cousins (1972); Eggen \& Phillips 
(unpublished); Lee et al. (1972) \nl
1980N  &  Hamuy et al. (1991) \nl
1981B  &  Barbon et al. (1982); Buta \& Turner (1983); Tsvetkov (1982) \nl
1986G  &  Phillips et al. (1987) \nl
1989B  &  Wells et al. (1994) \nl
1990N  &  Lira et al. (1998) \nl
1991T  &  Lira et al. (1998) \nl
1991bg &  Filippenko et al. (1992); Leibundgut et al. (1993) \nl
1992A  &  Suntzeff et al. (unpublished) \nl
1994D  &  Smith et al. (unpublished) \nl
1996X  &  Covarrubias et al. (unpublished) \nl
1998bu &  Suntzeff et al. (1999) \nl
\enddata
\end{deluxetable}

\clearpage

\begin{deluxetable}{lcccccc}
\tablenum{2}
\tablewidth{0pt}
\tablecaption{Host Galaxy Reddenings\label{t2}}
\tablehead{
\colhead{SN} &
\colhead{$\dmm_{obs}$\tablenotemark{a}} &
\colhead{E$(B-V)_{Gal}$\tablenotemark{b}} &
\colhead{E$(B-V)_{Tail}$\tablenotemark{b}} &
\colhead{E$(B-V)_{Max}$\tablenotemark{b}} &
\colhead{E$(V-I)_{Max}$\tablenotemark{b}} &
\colhead{E$(B-V)_{Avg}$\tablenotemark{a}}
}
\startdata
1937C &  0.87(10) & 0.014(001) &  0.012(053) &  0.061(060) &   \nodata &   0.03(03) \nl
1972E &  0.87(10) & 0.056(006) & -0.003(051) &  0.040(060) & -0.001(073) & 0.01(03) \nl
1980N &  1.28(04) & 0.021(002) &  0.110(052) &  0.077(044) & -0.009(056) & 0.05(02) \nl
1981B &  1.10(07) & 0.018(002) &  0.084(053) &  0.150(044) &   \nodata &   0.11(03) \nl
1986G &  1.73(07) & 0.115(011) &  0.668(057) &   \nodata   &   \nodata &   0.50(05) \nl
1989B &  1.31(07) & 0.032(003) &  0.457(067) &  0.363(069) &  0.465(077) & 0.34(04) \nl
1990N &  1.07(05) & 0.026(003) &  0.160(057) &  0.106(051) &  0.057(061) & 0.09(03) \nl
1990O &  0.96(10) & 0.093(009) &  0.069(060) &  0.072(061) & -0.071(073) & 0.02(03) \nl
1990T &  1.15(10) & 0.053(005) &  0.138(056) &  0.080(104) &  0.052(112) & 0.09(04) \nl
1990Y &  1.13(10) & 0.008(001) &  0.348(061) &  0.388(106) &  0.064(111) & 0.23(04) \nl
1990af & 1.56(05) & 0.035(004) & -0.008(111) &  0.052(047) &   \nodata &   0.04(03) \nl
1991S &  1.04(10) & 0.026(003) &  0.122(069) &  0.089(104) &  0.004(111) & 0.06(04) \nl
1991T &  0.94(05) & 0.022(002) &  0.174(052) &   \nodata   &   \nodata   & 0.14(05)\tablenotemark{d} \nl
1991U &  1.06(10) & 0.062(006) &  0.145(063) &  0.141(105) &  0.152(112) & 0.11(04) \nl
1991ag & 0.87(10) & 0.062(006) &  0.075(054) &  0.122(061) &  0.030(080) & 0.07(03) \nl
1991bg & 1.93(10) & 0.040(004) &  0.038(064) &   \nodata   &   \nodata &   0.03(05) \nl
1992A &  1.47(05) & 0.017(002) &  0.004(051) &  0.000(046) & -0.061(059) & 0.00(02) \nl
1992J &  1.56(10) & 0.057(006) &  0.050(065) &  0.129(106) & -0.064(114) & 0.03(04) \nl
1992K &  1.93(10) & 0.101(010) & -0.029(055) &  \nodata    &   \nodata &   0.00(04) \nl
1992P &  0.87(10) & 0.021(002) &  0.082(058) &  0.064(046) &  0.093(072) & 0.07(03) \nl
1992ae & 1.28(10) & 0.036(004) &  0.215(097) &  0.132(060) &   \nodata &   0.12(04) \nl
1992ag & 1.19(10) & 0.097(010) &  0.622(066) &  0.151(061) &  0.116(072) & 0.10(04)\tablenotemark{c} \nl
1992al & 1.11(05) & 0.034(003) &  0.052(053) & -0.015(044) & -0.013(068) & 0.01(03) \nl
1992aq & 1.46(10) & 0.012(001) &  0.137(096) &  0.115(061) & -0.200(074) & 0.00(04) \nl
1992au & 1.49(10) & 0.017(002) &  0.031(093) &  0.058(105) & -0.243(113) & 0.00(04) \nl
1992bc & 0.87(05) & 0.022(002) & -0.055(053) & -0.006(044) &  0.012(069) & 0.00(02) \nl
1992bg & 1.15(10) & 0.185(018) &  0.050(059) &  0.009(062) & -0.016(075) & 0.01(03) \nl
1992bh & 1.05(10) & 0.022(002) &  0.177(071) &  0.132(060) &  0.126(071) & 0.12(03) \nl
1992bk & 1.57(10) & 0.015(001) &  0.034(065) &  0.001(062) & -0.013(114) & 0.01(03) \nl
1992bl & 1.51(10) & 0.011(001) & -0.006(065) &  0.012(061) & -0.073(075) & 0.00(03) \nl
1992bo & 1.69(05) & 0.027(003) & -0.004(058) & -0.014(049) &  0.020(064) & 0.00(03) \nl
1992bp & 1.32(10) & 0.069(007) &  0.145(077) & -0.034(060) & -0.090(073) & 0.00(03) \nl
1992br & 1.69(10) & 0.026(003) & -0.002(102) &  0.016(063) &   \nodata &   0.01(04) \nl
1992bs & 1.13(10) & 0.012(001) &  0.142(102) &  0.123(060) &   \nodata &   0.10(04) \nl
1993B &  1.04(10) & 0.079(008) &  0.200(069) &  0.196(061) &  0.047(072) & 0.12(03) \nl
1993H &  1.69(10) & 0.060(006) &  0.064(057) &   \nodata   &   \nodata &   0.05(04) \nl
1993O &  1.22(05) & 0.053(005) &  0.059(059) & -0.046(044) &  0.025(069) & 0.00(03) \nl
1993ac & 1.19(10) & 0.163(016) &  0.080(089) &  0.096(062) &  0.250(074) & 0.12(04) \nl
1993ae & 1.43(10) & 0.039(004) & -0.024(053) & -0.036(104) & -0.009(113) & 0.00(03) \nl
1993ag & 1.32(10) & 0.112(011) &  0.152(061) &  0.112(061) & -0.025(074) & 0.07(03) \nl
1993ah & 1.30(10) & 0.020(002) &  0.075(074) & -0.013(104) & -0.062(112) & 0.01(04) \nl
1994D &  1.32(05) & 0.022(002) & -0.060(052) & -0.017(044) & -0.010(057) & 0.00(02) \nl
1994M &  1.44(10) & 0.024(002) &  0.115(064) &  0.057(061) &  0.126(074) & 0.08(03) \nl
1994Q &  1.03(10) & 0.017(002) &  0.090(065) &  0.099(104) &  0.048(111) & 0.06(04) \nl
1994S &  1.10(10) & 0.021(002) &  0.001(112) &  0.009(045) & -0.034(071) & 0.00(03) \nl
1994T &  1.39(10) & 0.029(003) &  0.105(114) &  0.126(061) &  0.112(073) & 0.09(04) \nl
1994ae & 0.86(05) & 0.031(003) &  0.184(056) &  0.175(045) &  0.053(057) & 0.12(03) \nl
1995D &  0.99(05) & 0.058(006) &  0.103(054) &  0.044(044) &  0.016(056) & 0.04(02) \nl
1995E &  1.06(05) & 0.027(003) &  0.934(072) &  0.793(045) &  0.992(068) & 0.74(03) \nl
1995ac & 0.91(05) & 0.042(004) &  0.109(054) &   \nodata   &   \nodata   & 0.08(04)\tablenotemark{d} \nl
1995ak & 1.26(10) & 0.043(004) &  0.260(057) &  0.107(060) &  0.294(072) & 0.18(03) \nl
1995al & 0.83(05) & 0.014(001) &  0.157(064) &  0.246(045) &   \nodata   & 0.15(03) \nl
1995bd & 0.84(05) & 0.495(050) &  0.242(074) &   \nodata   &   \nodata   & 0.15(06)\tablenotemark{d} \nl
1996C &  0.97(10) & 0.014(001) &  0.078(055) &  0.139(060) &  0.106(072) & 0.09(03) \nl
1996X &  1.25(05) & 0.069(007) & -0.010(051) & -0.016(044) &  0.076(056) & 0.01(02) \nl
1996Z &  1.22(10) & 0.063(006) &  0.500(073) &  0.352(061) &   \nodata &   0.33(04) \nl
1996ai & 0.99(10) & 0.014(001) &  2.015(080) &  1.936(065) &  1.575(072) & 1.44(04) \nl
1996bk & 1.75(10) & 0.018(002) &  0.274(064) &   \nodata   &   \nodata &   0.19(05) \nl
1996bl & 1.17(10) & 0.105(011) &  0.113(063) &  0.095(061) &  0.090(073) & 0.08(03) \nl
1996bo & 1.25(05) & 0.078(008) &  0.358(053) &  0.385(061) &  0.253(069) & 0.28(03) \nl
1998bu & 1.01(05) & 0.025(003) &  0.328(053) &  0.375(054) &  0.519(074) & 0.33(03) \nl
1996bv & 0.93(10) & 0.105(010) &  0.215(062) &  0.233(061) &  0.318(073) & 0.21(03) \nl
\enddata
\tablenotetext{a}{The mean errors are listed in units of 0\fm01.}
\tablenotetext{b}{The mean errors are listed in units of 0\fm001.}
\tablenotetext{c}{E$(B-V)_{Tail}$ measurement ignored in calculation of 
E$(B-V)_{Avg}$ due to liklihood of systematic errors in late-time photometry.}
\tablenotetext{d}{1991T-like event; E$(B-V)_{Avg}$ based only on value of
E$(B-V)_{Tail}$.}
\end{deluxetable}

\clearpage

\begin{deluxetable}{cccccc}
\tablenum{3}
\tablewidth{0pt}
\tablecaption{Fits to Decline Rate versus Luminosity Relation\tablenotemark{a}
$\Delta M_{max} = a[\dmm-1.1]+b[\dmm-1.1]^2$\label{t3}}
\tablehead{
\colhead{Bandpass} &
\colhead{a\tablenotemark{b}} &
\colhead{b\tablenotemark{b}} &
\colhead{$\sigma$(mag)} &
\colhead{$\chi^2_\nu$} &
\colhead{n}
}
\startdata
$B$ & 0.786(398) & 0.633(742) & 0.11 & 0.47 & 17 \nl     
$V$ & 0.672(396) & 0.633(742) & 0.09 & 0.42 & 17 \nl     
$I$ & 0.422(400) & 0.633(742) & 0.13 & 0.83 & 15 \nl     
\enddata
\tablenotetext{a}{$\Delta M_{max} = M_{max} - M_{max}(\dmm = 1.1)$}
\tablenotetext{b}{The errors are listed in units of 0\fm001.}
\end{deluxetable}

\newpage

\begin{figure}
\plotfiddle{Phillips.fig1.ps}{7in}{-90}{80}{80}{-310}{550}
{\center Phillips {\it et al.} Figure~\ref{f1}}
\end{figure}

\begin{figure}
\epsscale{0.8}
\plotone{Phillips.fig2.ps} 
{\center Phillips{\it et al.} Figure~\ref{f2}}
\end{figure}

\begin{figure}
\plotfiddle{Phillips.fig3.ps}{7in}{-90}{80}{80}{-310}{500}
{\center Phillips {\it et al.} Figure~\ref{f3}}
\end{figure}

\begin{figure}
\plotfiddle{Phillips.fig4.ps}{7in}{0}{80}{80}{-250}{-25}
{\center Phillips {\it et al.} Figure~\ref{f4}}
\end{figure}

\begin{figure}
\plotfiddle{Phillips.fig5.ps}{7in}{0}{80}{80}{-250}{-25}
{\center Phillips {\it et al.} Figure~\ref{f5}}
\end{figure}

\begin{figure}
\epsscale{0.8}
\plotone{Phillips.fig6.ps} 
{\center Phillips{\it et al.} Figure~\ref{f6}}
\end{figure}

\begin{figure}
\plotone{Phillips.fig7.ps} 
{\center Phillips{\it et al.} Figure~\ref{f7}}
\end{figure}

\begin{figure}
\plotfiddle{Phillips.fig8.ps}{7in}{-90}{75}{75}{-290}{500}
{\center Phillips {\it et al.} Figure~\ref{f8}}
\end{figure}

\begin{figure}
\plotfiddle{Phillips.fig9.ps}{7in}{0}{80}{80}{-250}{-25}
{\center Phillips {\it et al.} Figure~\ref{f9}}
\end{figure}

\begin{figure}
\plotfiddle{Phillips.fig10.ps}{7in}{0}{80}{80}{-250}{-25}
{\center Phillips {\it et al.} Figure~\ref{f10}}
\end{figure}

\end{document}